\documentstyle[12pt]{article}
\def\H{\mbox{I\hspace{-.15em}H\hspace{-.15em}I}}
\def\1{\mbox{I\hspace{-.15em}1}}

\def\Z{\mbox{Z\hspace{-.3em}Z}}

\def\R{{\rm I\hspace{-.15em}R}}

\def\b{\begin{equation}}
\def\e{\end{equation}}
\def\bee{\begin{enumerate}}
\def\eee{\end{enumerate}}

\title{N=1 de Sitter Supersymmetry Algebra}

\author{A. Pahlavan$^1$, S. Rouhani$^{1}$ and M.V.
Takook$^{2}$\thanks{e-mail: takook@razi.ac.ir}}
\date{\today}

\begin{document}

\maketitle {\it   \centerline{\it $^1 $ Plasma Physics Research
Center, Islamic Azad University,}
 \centerline{\it P.O.BOX 14835-157, Tehran,
IRAN} \centerline{$^2$  Department of Physics, Razi University,
Kermanshah, IRAN}}

\begin{abstract}

Recalling the universal covering group of de Sitter universe, the
transformation properties of the spinor fields $\psi(x)$ and
${\overline\psi}(x)$, in the ambient space notation, are presented
in this paper. The charge conjugation symmetry of the de Sitter
spinor field is then discussed in the above notation. Using this
spinor field and charge conjugation, de Sitter supersymmetry
algebra in the ambient space notation has been established. It is
shown that a novel de Sitter superalgebra can be attained by the
use of spinor field and charge conjugation in the ambient space
notation.

\end{abstract}

\vspace{0.5cm} {\it Proposed PACS numbers}: 04.62.+v, 98.80.Cq,
12.10.Dm \vspace{0.5cm}

\section{Introduction}

Recent astrophysical data received from type Ia supernovas
indicate that our universe might currently be in a de Sitter (dS)
phase. Therefore it is important to find a formulation of de
Sitter quantum field theory with the same level of completeness
and rigor as for its Minkowskian counterpart. Some questions,
however,  are usually put forth for the non-existence of
supersymmetry models with a positive cosmological constant, {\it
i.e.} supersymmetry in de Sitter space. Such arguments are often
based on the non-existence of Majorana spinors for $O(4,1)$
\cite{piniso,dwhe}. Pilch et al have shown that if for every
spinor, its independent charge-conjugate could be defined, de
Sitter supergravity can be established with even $N$
\cite{piniso}.

Bros et al. \cite{brgamo} presented a QFT of scalar free field in
de Sitter space that closely mimics QFT in Minkowski space. We
have generalized the Bros's quantization of field, to quantization
of fields with various spins in de Sitter space \cite{ta}. In
continuation of previous works where the charge-conjugate spinor
had been defined \cite{rota3,morota}, the supersymmetry in the
ambient space notation has been studied in the present paper.
Section two has been devoted to the discussion of the de Sitter
group $SO(1,4)$, {\it i.e.} space-time symmetry of de Sitter
space, and its universal covering group $Sp(2,2)$. Recalling the
transformation properties of the spinor fields
$\psi(x),{\overline\psi}(x)$, the charge conjugation symmetry of
the de Sitter spinor field in ambient space notation is discussed
in section $3$. The general de Sitter superalgebra is presented in
section $4$. It is shown that a novel dS-superalgebra can be
attained by the use of the spinor fields and charge conjugation in
the ambient space notation. Finally, a brief conclusion and an
outlook have been given in section $5$. To illustrate the novel
algebra, the generalized Jacobi identities are calculated in
appendix A.

\section{de Sitter group}

The de Sitter space is an elementary solution of the positive
cosmological Einstein equation in the vacuum. It is conveniently
seen as a hyperboloid embedded in a five-dimensional Minkowski
space
     \b X_H=\{x \in \R^5 ;x^2=\eta_{\alpha\beta} x^\alpha x^\beta
=-H^{-2}\},\;\;
      \alpha,\beta=0,1,2,3,4, \e
where $\eta_{\alpha\beta}=$diag$(1,-1,-1,-1,-1)$. The de Sitter
metrics reads
$$ds^2=\eta_{\alpha\beta}dx^{\alpha}dx^{\beta}\mid_{x^2=-H^{-2}}=g_{\mu\nu}^{dS}dX^{\mu}dX^{\nu},\;\; \mu=0,1,2,3,$$
where the $X^\mu$'s are the $4$ space-time intrinsic coordinates
on dS hyperboloid. Different coordinate systems can be chosen for
$X^\mu$. A $10$-parameter group $SO_0(1,4)$ is the kinematical
group of the de Sitter universe. In the limit $H=0$, this reduces
to the Poincar\'e group. There are two Casimir operators
    $$ Q^{(1)}=-\frac{1}{2}L_{\alpha\beta} L^{\alpha\beta} ,$$
     \b
Q^{(2)}=-W_{\alpha}W^{\alpha},\;\;\;W_{\alpha}=-\frac{1}{8}\epsilon_
  {\alpha\beta\gamma\delta\eta}L^{\beta\gamma} L^{\delta\eta}, \e
where $\epsilon_ {\alpha\beta\gamma\delta\eta}$ is the usual
antisymmetrical tensor and the $L_{\alpha\beta}$'s are the
infinitesimal generators, which obey the commutation relations
\begin{equation} \lbrack L_{\alpha\beta}, L_{\gamma\delta}\rbrack
= -i(\eta_{\alpha\gamma}L_{\beta\delta}+\eta_{\beta\delta}
L_{\alpha\gamma}-\eta_{\alpha\delta}L_{\beta\gamma}-\eta_{\beta\gamma}
L_{\alpha\delta}).\label{eqcommf}
\end{equation} $L_{\alpha\beta}$
can be represented as
$L_{\alpha\beta}=M_{\alpha\beta}+S_{\alpha\beta}$, where $
M_{\alpha\beta}=-i(x_{\alpha}\partial_{\beta}-x_{\beta}\partial_{\alpha})$
is the ``orbital" part and $S_{\alpha\beta}$ is  the ``spinorial"
part. The form of the $S_{\alpha\beta}$ depends on the spin of the
field. For spin $\frac{1}{2}$ field, it can be defined as
\begin{equation}
S_{\alpha\beta}=-{i\over
4}\lbrack\gamma_{\alpha},\gamma_{\beta}\rbrack, \label{genspin}
\end{equation}
where the five  $4\times 4$ matrices $\gamma^{\alpha}$ are the
generators of the Clifford algebra based on the metric
$\eta_{\alpha\beta}$:
\begin{equation}
\gamma^{\alpha}\gamma^{\beta}+\gamma^{\beta}\gamma^{\alpha}=2\eta^{\alpha\beta}{\1}\,
,
\quad{\gamma^{\alpha}}^{\dag}=\gamma^{0}\gamma^{\alpha}\gamma^{0}.\label{Clifford}
\end{equation}
An explicit and convenient representation is provided by
\cite{ta1,brgamota,taka}
$$ \gamma^0=\left( \begin{array}{clcr} \1 & \;\;0 \\ 0 &-\1 \\ \end{array} \right)
 ,\gamma^4=\left( \begin{array}{clcr} 0 & \1 \\ -\1 &0 \\ \end{array} \right),  $$
  \b   \gamma^1=\left( \begin{array}{clcr} 0 & i\sigma^1 \\ i\sigma^1 &0 \\    \end{array} \right)
     ,\gamma^2=\left( \begin{array}{clcr} 0 & -i\sigma^2 \\ -i\sigma^2 &0 \\  \end{array} \right)
      , \gamma^3=\left( \begin{array}{clcr} 0 & i\sigma^3 \\ i\sigma^3 &0 \\   \end{array} \right),\e
where $ \1 $ is the unit $ 2\times2 $ matrix and $\sigma^i $ are
the Pauli matrices. This representation had been proved to be
useful in analysis of the physical relevance of the group
representation \cite{brgamota}. In this representation,
$$\quad{\gamma^{\alpha}}^{T}=\gamma^{4}\gamma^{2}\gamma^{\alpha}\gamma^{2}\gamma^{4}.$$

The spinor wave equation in de Sitter space-time has been
originally deduced by Dirac in 1935 \cite{dir}, and can be
obtained from the eigenvalue equation of the second order Casimir
operator \cite{ta1,brgamota}
\begin{equation}  (-i\not x\gamma.\bar{\partial} +2i+\nu)\psi(x)=0,
\end{equation}
where $\not x=\eta_{\alpha \beta} \gamma^\alpha x^\beta$ and
$\bar{\partial_{\alpha}}=\partial_{\alpha}+H^2x_{\alpha}x\cdot\partial$.
Due to covariance of the de Sitter group, the adjoint spinor is
defined as follows \cite{brgamota}:
\begin{equation} {\overline \psi}(x)\equiv
\psi^{\dag}(x){\gamma^0}{\gamma^4}.\label{adj} \end{equation}

Let us now recall the transformation properties of the spinor
fields $\psi(x)$ and ${\overline\psi}(x)$. The two-fold, universal
covering group of $SO_0(1,4)$, is the (pseudo-)symplectic group
$Sp(2,2)$, \cite{taka}
\begin{equation}
Sp(2,2)=\bigl\{g\in {\rm Mat}(2;{\H}) :\ \det g=1,\
g^{\dag}\gamma^{0}g=\gamma^{0} \bigr\}, \end{equation} where
$g^{\dag}=^T\!\!\tilde{g}$, $^T\!\!g$ is the transposed of $g$ and
$\tilde g$ the quaternionic conjugate of $g$. It should be noted
that quaternionic ${\cal P}$ is \b{\cal P}=(x^4 ,\vec x)=x^4
\1+ix^1\sigma^1-ix^2 \sigma^2+ix^3\sigma^3,\e where $\sigma^i$ are
the Pauli matrices and $ \tilde {\cal P }=(x^4 ,-\vec x) $ is its
conjugate. For obtaining the isomorphic relation between the two
groups we define the matrices $X$ associated with $x \in X_H$ by:
            \b   X=\left(\begin{array}{clcr} x^0 & \cal P \\  \tilde {\cal P } &  x^0 \\ \end{array}
           \right).\e
Through representation $(6)$ of the $\gamma$ matrices, $X$ can be
written in the following form:
             \b \not x=x.\gamma=X\gamma^0=\left(\begin{array}{clcr} x^0 &
            -\cal P \\  \tilde {\cal P } & - x^0 \\ \end{array} \right) .\e
The transformation of $X$, under the action of the group $Sp(2,2)$
is
    \b X'=gX\tilde g^{t} ,\;\;   \not x'  =g\not xg^{-1}.\e
For $\Lambda \in SO_0(1,4)$ and $g\in Sp(2,2)$ we have
           $$ x'^{\alpha}=\eta^{\alpha\beta}x'_{\beta}=\frac{1}{4}
         tr(\gamma^{\alpha}\gamma^{\beta})x'_{\beta}=
           \frac{1}{4}tr(\gamma^{\alpha}g\not xg^{-1})$$
           \b
         =\frac{1}{4}tr(\gamma^{\alpha}g\gamma^{\beta}g^{-1})x_{\beta}
          =\Lambda^{\alpha\beta}(g)x_{\beta}.\e
For every $g\in Sp(2,2)$, there exists a transformation
 $\Lambda \in SO_0(1,4)$, which satisfies the following relations \b
\Lambda^{\alpha}_{\beta}(g)=\frac{1}{4}tr(\gamma^{\alpha}
g\gamma_{\beta}g^{-1}),\;\;\;
\Lambda^{\alpha}_{\beta}\gamma^{\beta}=g\gamma^{\alpha}g^{-1}.
\label{isom} \e The isomorphic relation between the two groups is
\begin{equation}
SO_{0}(1,4)\approx Sp(2,2)/{{\Z}_{2}}\,.
\end{equation}
The transformation laws for the $\psi(x)$ and its adjoint
${\overline \psi}(x)$, under which the de Sitter-Dirac equation is
covariant, are :
\begin{eqnarray}
&& \psi(x)\  \rightarrow \
\psi^{\prime}(x)=g^{-1}\psi\bigr(\Lambda(g)x\bigr), \label{lawI}\\
&&{\overline\psi}(x)  \ \rightarrow\ {\overline
\psi}^{\prime}(x)={\overline
\psi}\bigr(\Lambda(g)x\bigr)i(g),\label{lawII}
\end{eqnarray}
where $i(g)\equiv-\gamma^{4}g\gamma^{4}$ \cite{brgamota,taka}.
Similar to the Minkowskian space, it is useful to define $g$ by \b
g=\exp[-\frac{i}{2}\omega^{\alpha\beta} S_{\alpha\beta}],\;\;\;\;
\omega^{\alpha\beta}=-\omega^{\beta\alpha},\e where $\gamma^0
g^\dag\gamma^0=g^{-1}$, {\it i.e.} $g \in Sp(2,2).$

\section{Charge conjugation}

Previously, the charge conjugation spinor $\psi^c$ was calculated
in the ambient space notation \cite{morota} \b \psi^{c}=\eta_c
C(\gamma^4)^T(\bar \psi)^T,\e where $\eta_c$ is an arbitrary
unobservable phase value, generally set to unity. In the present
framework charge conjugation is an antilinear transformation. In
the $\gamma$ representation $(6)$ we had obtained \cite{morota}:
$$ C\gamma^{0}C^{-1}=-\gamma^0 , C\gamma^4C^{-1}=-\gamma^4$$ \b
C\gamma^{1}C^{-1}=-\gamma^1,C\gamma^{3}C^{-1}=-\gamma^3,C\gamma^{2}C^{-1}=\gamma^2.
\e In this representation, $C$ commutes with $\gamma^2$, and
anticommutes with other $\gamma$-matrices. Therefore the simple
choice may be taken to be $C=\gamma^2$, where the following
relation is satisfied \b C=-C^{-1}=-C^T=-C^\dag.\e This clearly
illustrates the non-singularity of $C$.

The adjoint spinor, defined by ${\overline \psi}(x)\equiv
\psi^{\dag}(x){\gamma^0}{\gamma^4}$, transforms in a different way
from $\psi$, under de Sitter transformation. It is easily shown
that $\psi^c$ transforms in the same way as $\psi$ \cite{morota}
$$\psi'^c(x')=g(\Lambda)\psi^c(x).$$ The wave equation of $\psi^c$
is different from the wave equation of $\psi$ by the sign of the
''charge'' $q$ and $\nu$ \cite{morota}. Thus it follows that if
$\psi$ describes the motion of a dS-Dirac ''particle'' with the
charge $q$, $\psi^c$ represents the motion of a dS-Dirac
''anti-particle'' with the charge $(-q)$. In other words $\psi$
and ${\psi}^c$ can describe "particle'' and "antiparticle'' wave
functions. $\psi$ and ${\psi}^c$ are charge conjugation of each
other \b ({\psi}^c)^c=C\gamma^0{{\psi}^c}^\ast=C\gamma^0(C
\gamma^0\psi)=\psi. \e

\section{N=1 Supersymmetry Algebra}

Supersymmetry in de Sitter space has been studied by Pilch et al
\cite{piniso}. Recently, supersymmetry has been investigated in
constant curvature space as well \cite{mcsh,fell}. In this section
we have presented the supersymmetry algebra in ambient space
notation. It is shown that if the spinor field and the charge
conjugation operators are defined in the ambient space notation, a
novel de Sitter superalgebra can be attained.

In order to extend the de Sitter group, the generators of
supersymmetry transformation $Q^n_i$ are introduced. Here $i$ is
the spinor index ($i=1,2,3,4$) and $n$ is the supersymmetry index
$ n=1,..,N$. $Q^n_i$'s are superalgebra spinor generators which
transform as de Sitter group spinors. The generators
$\tilde{Q}^n_i$ are defined by \b \tilde{Q}_i=\left(Q^T \gamma^4
C\right)_i=\bar{Q^c}_i,\e where $Q^T$ is the transpose of $Q$. It
can be shown that $\tilde{Q}\gamma^4 Q$ is a scalar field under
the de Sitter transformation \cite{morota}.

For $N\neq 1$, closure of algebra requires extra bosonic
generators. These do not necessarily commute with other generators
and consequently are not central charges. They are internal
symmetry generators \cite{piniso}. These generators, shown by
$T_{mn}$, commute with de Sitter generators.

Therefore the de Sitter superalgebra in four-dimensional
space-time has the following generators:
\begin{itemize}
    \item $M_{\alpha\beta}$, the generators of de Sitter group,
    \item the internal group generators $T_{nm}$, that are defined by
    the additional condition
$$T_{nm}=-T_{mn}; n,m=1,2....N,$$
    \item the $4$-component dS-Dirac spinor generators,
$$Q^n_i,\;\;\;i=1,2,3,4,\;\;\; n=1,2,...,N.$$
 \end{itemize}

To every generator $A$, a grade $p_a$ is assigned. For the
fermionic generator $p_a=1$, and for the bosonic generator
$p_a=0$. The bilinear product operator is defined by \b
[A,B]=-(-1)^{p_a.p_b} [B,A].\e The generalized Jacobi identities
is \b (-1)^{p_a.p_c} [A,[B,C]]+(-1)^{p_c.p_b}[C,
[A,B]]+(-1)^{p_b.p_a} [B,[C,A]]=0.\e Using different generalized
Jacobi identities, similar to the method presented by Pilch el al.
\cite{piniso}, the full dS-superalgebra can be written in the
following form:
$$ [M_{\alpha\beta}, M_{\gamma\delta}] =
-i(\eta_{\alpha\gamma}M_{\beta\delta}+\eta_{\beta\delta}
M_{\alpha\gamma}-\eta_{\alpha\delta}M_{\beta\gamma}-\eta_{\beta\gamma}
M_{\alpha\delta}),$$ $$
[T_{rl},T_{pm}]=-i(\omega_{rp}T_{lm}+\omega_{lm}T_{rp}-\omega_{rm}T_{lp}-\omega_{lp}T_{rm}),$$
$$ [M_{\alpha\beta}, T_{rl}] =0,$$
$$[Q^{r}_{i},M_{\alpha\beta}]=(S_{\alpha\beta}Q^{r})_{i},\;\;
[\tilde Q^{r}_{i},M_{\alpha\beta}]=-(\tilde Q^{r}
S_{\alpha\beta})_{i}$$
$${[Q^{r}_{i},T^{lp}]}={-(\omega^{rl}Q^{p}_{i}-\omega^{rp}Q^{l}_{i})}$$
$$ \{Q^{r}_{i},Q^{l}_{j}\}=\omega^{rl}\left(S^{\alpha\beta}\gamma^4
\gamma^2\right)_{ij}M_{\alpha\beta}+ \left(\gamma^4
\gamma^2\right)_{ij}T^{rl},$$ where $S_{\alpha\beta}$ is defined
by $(4)$. The following relations are used to determine the above
structure \b \left(S^{\alpha\beta}\gamma^4 \gamma^2\right)^T
=\left(S^{\alpha\beta}\gamma^4
\gamma^2\right),\;\;\left(\gamma^\alpha \gamma^4 \gamma^2\right)^T
=-\gamma^\alpha \gamma^4 \gamma^2,\;\; \left(\gamma^4
\gamma^2\right)^T =-\gamma^4 \gamma^2.\e It is necessary to obtain
matrix $\omega$, which determines the structure of the internal
group. For even $N$ de Sitter supersymmetry algebra, studied by
Pilch et al \cite{piniso}, the matrix $\omega$ has been obtained
explicitly. A new dS-superalgebra, defined in the ambient space
notation, is introduced in this stage for $N=1$ case. In this
case, $T_{11}=0, \;\omega=1 $ and the simple de Sitter
supersymmetry algebra is defined by the following relations: \b
[M_{\alpha\beta}, M_{\gamma\delta}] =
-i(\eta_{\alpha\gamma}M_{\beta\delta}+\eta_{\beta\delta}
M_{\alpha\gamma}-\eta_{\alpha\delta}M_{\beta\gamma}-\eta_{\beta\gamma}
M_{\alpha\delta}),\e \b \{Q_i,Q_j\}=\left(S^{\alpha\beta}\gamma^4
\gamma^2\right)_{ij}M_{\alpha\beta},\e \b
[Q_i,M_{\alpha\beta}]=(S_{\alpha\beta}Q)_i, \;\; [\tilde
Q_i,M_{\alpha\beta}]=-(\tilde Q S_{\alpha\beta})_i.\e This can be
proved by the use of generalized Jacobi identities (appendix).

Finally we present an explicit form of the supersymmetric
generators which satisfy the above relations. We consider a
superspace with bosonic coordinates $x^\alpha$ and fermionic
coordinates $\theta_i$ where $\theta_i$ is a four component de
Sitter-Dirac Grassmann spinor in the ambient space notation. The
suitable representation of
these superalgebra generators in superspace are provided by \b \left\{%
\begin{array}{ll}
    M_{\alpha\beta}=-i(x_{\alpha}\bar \partial_{\beta}-x_{\beta}\bar \partial_{\alpha})+\frac{\partial}{\partial \theta}S_{\alpha\beta}\theta, \\
    Q=\gamma.\bar{\partial} \theta+ \not x\frac{\partial}{\partial \tilde\theta},   \\
\end{array}%
\right. \e where $\frac{\partial}{\partial
\tilde\theta_i}=\left(\gamma^2
\gamma^4\right)_{ik}\frac{\partial}{\partial \theta_k}$. Using the
equation $\{ \theta_i, \frac{\partial}{\partial
\theta_j}\}=\delta_{ij}$ and the following identities \cite{mcsh},
$$ (S^{\alpha\beta}\gamma^4
\gamma^2)_{ij}(S_{\alpha\beta})_{kl}+\left(S^{\alpha\beta}\gamma^4
\gamma^2\right)_{jk}(S_{\alpha\beta})_{il}+\left(S^{\alpha\beta}\gamma^4
\gamma^2\right)_{ki}(S_{\alpha\beta})_{jl}=0,$$
$$(S^{\alpha\beta})_{ij}(S_{\alpha\beta})_{kl}+(S^{\alpha\beta})_{il}(S_{\alpha\beta})_{kj}=
(\gamma^{\alpha})_{ij}(\gamma_{\alpha})_{kl}+(\gamma^{\alpha})_{il}(\gamma_{\alpha})_{kj},$$
it is straightforward to prove the above de Sitter supersymmetry
algebra.

\section{ Conclusions}

The formalism of the quantum field in de Sitter universe, in
ambient space notation, is very similar to the quantum field
formalism in Minkowski space. In this paper we present the de
Sitter supersymmetry algebra in this notation, which is
independent of the choice of the coordinate system. In addition, a
novel superalgebra $(28-30)$ has been obtained, which do not fall
into the categories considered in previous works
\cite{piniso,mcsh,na}. The importance of this formalism may be
shown further by the consideration of the linear quantum gravity
and supergravity in de Sitter space, which lays a firm ground for
further study of the evolution of the universe.

\vspace{0.5cm} \noindent {\bf{Acknowledgements}}: The authors
would like to extend their gratitude to R. Kallosh for the useful
discussions and S. Moradi for interest in this work, and the
referee for his constructive suggestions. \vspace{0.5cm}

\begin{appendix}

\section{$N=1$ de Sitter superalgebra}

The generalized Jacobi identities $(M,M,Q)$ and $(M,Q,Q)$ can be
easily utilized to prove the de Sitter supersymmetry algebra
$(28-30)$. The generalized Jacobi identity $(M,M,Q)$ is \b
[[M_{\alpha\beta},M_{\gamma\delta}],Q_{i}]+[[Q_{i},M_{\alpha\beta}],
M_{\gamma\delta}]+[[M_{\gamma\delta},Q_{i}],M_{\alpha\beta}]=0.\e
Using the equations $(28)$ and $(30)$, the following relations can
be obtained:
$$
[[M_{\alpha\beta},M_{\gamma\delta}],Q_{i}]=-i\eta_{\alpha\gamma}[M_{\beta\delta},Q_{i}]-
i\eta_{\beta\delta}[M_{\alpha\gamma},Q_{i}]
+i\eta_{\alpha\delta}[M_{\beta\gamma},Q_{i}]+i\eta_{\beta\gamma}[M_{\alpha\delta},Q_{i}],$$
$$
[[Q_{i},M_{\alpha\beta}],M_{\gamma\delta}]=[(S_{\alpha\beta})^{j}_{i}Q_{j},M_{\gamma\delta}]
=(S_{\alpha\beta})^{j}_{i}(S_{\gamma\delta})^{k}_{j}Q_{k},$$ $$
[[M_{\gamma\delta},Q_{i}],M_{\alpha\beta}]=[-(S_{\gamma\delta})^{j}_{i}Q_{j},M_{\alpha\beta}]
=-(S_{\gamma\delta})^{j}_{i}(S_{\alpha\beta})^{k}_{j}Q_{k}.$$
Implementing the above equations in the eq $(32)$,
$$-i\eta_{\alpha\gamma}[M_{\beta\delta},Q_{i}]-i\eta_{\beta\delta}[M_{\alpha\gamma},Q_{i}]
+i\eta_{\alpha\delta}[M_{\beta\gamma},Q_{i}]+i\eta_{\beta\gamma}[M_{\alpha\delta},
Q_{i}]+(S_{\alpha\beta})^{j}_{i}(S_{\gamma\delta})^{k}_{j}Q_{k}
-(S_{\gamma\delta})^{j}_{i}(S_{\alpha\beta})^{k}_{j}Q_{k}=0,$$
$$
i\eta_{\alpha\gamma}(S_{\beta\delta})^{j}_{i}Q_{j}+i\eta_{\beta\delta}(S_{\alpha\gamma})^{j}_{i}Q_{j}
-i\eta_{\alpha\delta}(S_{\beta\gamma})^{j}_{i}Q_{j}-i\eta_{\beta\gamma}(S_{\alpha\delta})^{j}_{i}Q_{j}
+[(S_{\alpha\beta})^{j}_{i}(S_{\gamma\delta})^{k}_{j}-(S_{\gamma\delta})^{j}_{i}(S_{\alpha\beta})^{k}_{j}]Q_{k}=0,$$
$$
i[\eta_{\alpha\gamma}(S_{\beta\delta}Q)_{i}+\eta_{\beta\delta}(S_{\alpha\gamma}Q)_{i}
-\eta_{\alpha\delta}(S_{\beta\gamma}Q)_{i}-\eta_{\beta\gamma}
(S_{\alpha\delta}Q)_{i}]+[(S_{\alpha\beta}S_{\gamma\delta}-S_{\gamma\delta}S_{\alpha\beta})Q]_{i}=0,$$
$$
i[(\eta_{\alpha\gamma}S_{\beta\delta}+\eta_{\beta\delta}S_{\alpha\gamma}
-\eta_{\alpha\delta}S_{\beta\gamma}-\eta_{\beta\gamma}S_{\alpha\delta})Q]_{i}+
[(S_{\alpha\beta}S_{\gamma\delta}-S_{\gamma\delta}S_{\alpha\beta})Q]_{i}=0,$$
it is shown that $S_{\alpha\beta}$'s satisfy the following
commutation relation $$ [S_{\alpha\beta},S_{\gamma\delta}]=-
i(\eta_{\alpha\gamma}S_{\beta\delta}+\eta_{\beta\delta}S_{\alpha\gamma}
-\eta_{\alpha\delta}S_{\beta\gamma}-\eta_{\beta\gamma}S_{\alpha\delta}).$$
This is none other than equation $(3$). This generalized Jacobi
identity verifies the relation $(30)$, {\it i.e.} commutation
relation of $Q$ and $M$.

The generalized Jacobi identity $(M,Q, Q)$ is \b
[\{{Q_{i},Q_{j}}\},M_{\alpha\beta}]+\{[M_{\alpha\beta},Q_{i}],
Q_{j}\}-\{[Q_{j},M_{\alpha\beta}],Q_{i}\}=0.\e By the use of
equations $(29)$ and $(30)$, we obtain $$
\left(S^{\gamma\delta}\gamma^4
\gamma^2\right)_{ij}\left[M_{\gamma\delta},M_{\alpha\beta}\right]-\left(S_{\alpha\beta}\right)_{ik}\{Q_k,
Q_{j}\}-\left(S_{\alpha\beta}\right)_{ik}\{ Q_k ,Q_{i}\}=0,$$
$$ \left(S^{\gamma\delta}\gamma^4
\gamma^2\right)_{ij}\left[M_{\gamma\delta},M_{\alpha\beta}\right]-\left(S_{\alpha\beta}\right)_{ik}\left(S^{\gamma\delta}\gamma^4
\gamma^2\right)_{kj}M_{\gamma\delta}-\left(S_{\alpha\beta}\right)_{jk}\left(S^{\gamma\delta}\gamma^4
\gamma^2\right)_{ki}M_{\gamma\delta}=0,$$
$$ \left(S^{\gamma\delta}\gamma^4
\gamma^2\right)_{ij}\left[M_{\gamma\delta},M_{\alpha\beta}\right]-\left(S_{\alpha\beta}
S^{\gamma\delta}\gamma^4
\gamma^2\right)_{ij}M_{\gamma\delta}-\left(S_{\alpha\beta}S^{\gamma\delta}\gamma^4
\gamma^2\right)_{ji}M_{\gamma\delta}=0.$$ Utilizing equation
$(27)$, we obtain $$ \left(S^{\gamma\delta}\gamma^4
\gamma^2\right)_{ij}\left[M_{\gamma\delta},M_{\alpha\beta}\right]-\left(S_{\alpha\beta}
S^{\gamma\delta}\gamma^4
\gamma^2\right)_{ij}M_{\gamma\delta}-\left(S^{\gamma\delta}S_{\alpha\beta}\gamma^2\gamma^4
\right)_{ij}M_{\gamma\delta}=0.,$$   $$
S^{\gamma\delta}\left[M_{\gamma\delta},M_{\alpha\beta}\right]-\left[S_{\alpha\beta},
S^{\gamma\delta}\right]M_{\gamma\delta}=0.$$ This generalized
Jacobi identity once again, verifies the anti-commutation relation
of $Q_i$ and $Q_j$.

\end{appendix}

\end{document}